# Deep learning-based instance segmentation for the precise automated quantification of digital breast cancer immunohistochemistry images

Blanca Maria Priego-Torres[a,b,*], Barbara Lobato-Delgado[b], Lidia Atienza-Cuevas[b,c], Daniel Sanchez-Morillo[a,b]

[a]*Biomedical Engineering and Telemedicine Research Group, Department of Automation Engineering, Electronics and Computer Architecture and Networks, School of Engineering, University of Cádiz, Avda. Universidad de Cádiz 10, Puerto Real, Cádiz, 11510, Spain*
[b]*Biomedical Research and Innovation Institute of Cadiz (INiBICA), Hospital Universitario Puerta del Mar, Avda Ana de Viya 21, Cádiz, 11009, Spain*
[c]*Department of Pathology, Hospital Universitario Puerta del Mar, Avda. Ana de Viya 21, Cádiz, 11009, Spain*

## Abstract

The quantification of biomarkers on immunohistochemistry breast cancer images is essential for defining appropriate therapy for breast cancer patients, as well as for extracting relevant information on disease prognosis. This is an arduous and time-consuming task that may introduce a bias in the results due to intra- and inter-observer variability which could be alleviated by making use of automatic quantification tools. However, this is not a simple processing task given the heterogeneity of breast tumors that results in non-uniformly distributed tumor cells exhibiting different staining colors and intensity, size, shape, and texture, of the nucleus, cytoplasm and membrane.

In this research work we demonstrate the feasibility of using a deep learning-based instance segmentation architecture for the automatic quantification of both nuclear and membrane biomarkers applied to IHC-stained slides. We have solved the cumbersome task of training set generation with the design and implementation of a web platform, which has served as a hub for communication and feedback between researchers and pathologists as well as a system for the validation of the automatic image processing models. Through this tool, we have collected annotations over samples of HE, ER and Ki-67 (nuclear biomarkers) and HER2 (membrane biomarker) IHC-stained images. Using the same deep learning network architecture, we have trained two models, so-called nuclei- and membrane-aware segmentation models, which, once successfully validated, have revealed to be a promising method to segment nuclei instances in IHC-stained images. The quantification method proposed in this work has been integrated into the developed web platform and is currently being used as a decision support tool by pathologists.

## 1. Introduction

Breast carcinoma is one of the most common malignancies with the highest mortality rate among women in industrialized countries (Ferlay et al., 2019). Due to the aggressive behavior of some subtypes and given that the breast is an accessible organ for early diagnosis, breast cancer is a permanent object of study concerning diagnostic methods and treatment.

*Corresponding author. Tel.: +34 956015144; e-mail: blanca.priego@uca.es
*Email addresses:* blanca.priego@uca.es (Blanca Maria Priego-Torres), barbara.lobato@inibica.es (Barbara Lobato-Delgado), lidia.atienza.sspa@juntadeandalucia.es (Lidia Atienza-Cuevas), daniel.morillo@uca.es (Daniel Sanchez-Morillo)

To determine the diagnosis of the disease in breast cancer, some classical clinicopathological features derived from the histological analysis of primary breast cancer samples are used. These features include, among others, tumor size, histological type of tumor, cellular and nuclear pleomorphism, mitotic index and presence of necrosis or vascular invasion. However, these parameters on their own are not sufficient to determine a precise prognosis and predictive factors of this complex disease (Zaha, 2014). For this reason, several ancillary techniques, including immunohistochemistry (IHC) and molecular studies, are often used to guide treatment decisions, classify breast cancer into biologically distinct subtypes with different behaviors, and ultimately, serve as prognostic and predictive indicators.

IHC is a general term that covers many methods used to determine tissue constituents (the antigens) with the employment of specific antibodies that can be visualized through staining (De Matos et al., 2010). The detection of antigen-antibody interaction under an optical microscope can be achieved by labeling the antibody with a visual substance, which is combined with a fluorescent or, more frequently, chromogen label, and then performing colorimetric evaluation.

Different pathology guidelines (Duffy et al., 2017; Calvo et al., 2018; Burstein et al., 2019) recommend determining in all cases of breast cancer, in addition to histologic grade, several tumor IHC biomarkers. Particularly, to evaluate the prognosis and establish therapeutic options, the updated guidelines from the European Group on Tumor Markers (Duffy et al., 2017) specify as mandatory the measurement of estrogen (ER)-alpha receptors, progesterone (PR) receptors and human epidermal growth factor receptor 2 (HER2) for all patients with invasive breast cancer, as well as the quantification of proliferation marker Ki-67 for determining prognosis, especially if values are low or high. In short, ER, PR and Ki-67 are nuclear immunohistochemical markers with varying grouping complexity and their quantification requires counting the number of immunonegative (blue stain due to hematoxylin counterstain) and immunopositive (brown stain in bright field microscopy) tumor cells in given regions (López et al., 2008). On the other hand, the criteria for assessing the status of HER2 are based on the intensity and completeness of cell membrane immunostaining and the percentage of membrane-positive cells (Qaiser et al., 2018).

Despite the importance of an accurate evaluation of these biomarkers, their quantification depends on the subjective evaluation of staining color and intensity by a trained pathologist. This quantification or scoring is a time-consuming process in which errors are introduced due to intra-observer (variations in a single observer's interpretation of results) and inter-observer (subjective differences in interpretation between observers) variations (Kirkegaard et al., 2006). The quantification process usually involves the selection of hot-spot areas or regions of interest. Then, the staining of hundreds of cells that appear within the selected areas must be visually evaluated to score the IHC-stained preparation, which is cumbersome and error-prone. The subjectivity involved in these two steps makes the inter- and intra-observer variations of the scoring process not negligible, as has been already demonstrated (Leung et al., 2019).

Recently, huge advances in image acquisition devices have enabled histology technicians to scan conventional glass slides to produce high-quality digital slides, also known as whole slide images (WSI). This leads to pathologists moving from viewing glass slides in the microscope to navigating in a digital virtual slide similarly to how one can do in Google Maps (Zarella et al., 2019). It brings many new opportunities that cannot be achieved with traditional microscopes, including digital collaboration, working from remote sites, integration with electronic workflows and, what is relevant in connection with this work, the application of computer-aided diagnosis/prognosis (CAD) support tools based on artificial intelligence computing methods (Farahani et al., 2015). CAD tools are essential in the extension and establishment of digital pathology, given the urgent need to develop systems that support pathologists in their routine tasks, alleviating their workload and addressing issues related to the low reproducibility of diagnostic results.

Regarding the automatic scoring of IHC-stained images through automatic methods, there are a variety of commercial software that include tools designed for quantitative image analysis. Some examples are ACIS (ChromaVision Medical Systems, Inc., San Juan Capistrano, CA, USA), AQUA (HistoRx, New Haven, CT, USA), Ariol SL-50 (Applied Imaging, San Jose, CA, USA), BLISS and IHC score (Bacus Laboratories, Inc, Lombard, IL, USA), iVision and GenoMx (BioGenex, San Ramon, CA, USA), LSC Laser Scanning Cytometer (CompuCyte, Cambridge, MA, USA), ScanScope (Aperio Technologies, Inc., Vista, CA, USA), SlidePath's Tissue Image Analysis (Leica Biosystems, Wetzlar, Germany) and Virtuoso (Ventana Medical Systems, Tucson, AZ, USA) (Rojo et al., 2009). Several of these commercial applications have demonstrated more reproducible and uniform results than manual evaluation and have received approval for diagnostic use by the FDA (US Food and Drug Administration) and CE-Mark for In-Vitro Diagnosis (Garcia-Rojo et al., 2019). However, the majority of the mentioned software relies on conventional image processing techniques, based on the detection of hue, saturation and brightness levels (Chlipala et al., 2020), and some even implying the need for the pathologist to establish thresholds prior to processing. Likewise, multiple IHC quantification works based on conventional computer vision techniques, such as the implementation of morphological transformation schemes (Huang & Lai, 2010), modified watershed algorithms, (Shu et al., 2013; Akakin et al., 2012), local thresholding (CLT) method (Shu et al., 2020) and the spatial color algorithm (SCA) prior to thresholding (Barricelli et al., 2019) for nucleus region detection, can be found in the literature, exhibiting similar limitations. In this regard, new methods for the accurate quantification of IHC-stained images require, on the one hand, to be robust to non-uniformities that may appear between WSI, such as different staining intensity between different labs, background staining, tissue folding, etc. On the other hand, the new approaches need to take into account contextual information, i.e., not only distinguishing pixels according to hue, saturation and brightness levels, but also considering whether the pixels are part of tumor/non tumoral cells, artifacts or other structures that should be ignored in the quantification process. This involves the application of techniques capable of abstracting information in a more complex way, and this is where machine learning can offer its great potential.

As evidence thereof, the last decade has seen an increase in research into machine learning techniques applied to the quantification of digital breast cancer immunohistochemistry images analysis (Irshad et al., 2013). Generally speaking, these machine learning algorithms can be classified into hand-crafted and non-hand-crafted algorithms (Badejo et al., 2018). The former comprise those methods in which a specialist manually decides which image features are relevant to solve the processing tasks involved in the automatic quantification process. Examples of hand-crafted algorithms applied to IHC images rely on K-means clustering (Al-Lahham et al., 2012), support vector machines (SVM) (Chen et al., 2019a; Markiewicz et al., 2009), and online sparse dictionary learning methods (Xing et al., 2013). The latter, of which the most representative are the deep-learning techniques, learn these characteristics from the data automatically and efficiently, revealing a greater capacity for generalization. First attempts to use deep learning for the quantification of nuclear biomarkers (Saha et al., 2017; Sheikhzadeh et al., 2018; Narayanan et al., 2018), specifically Ki-67, and membrane biomarkers (Vandenberghe et al., 2017), addressed the problem in two steps, first extracting small patches in which the different nuclei appear centered in the image, and then classifying these into immunopositive or immunonegative cells through a deep learning model, which entails a high computational cost of classifying each nucleus independently and the inability to distinguish between tumor and non-tumor cells. Xue et al. (2016) employed a deep learning model to analyze the cell counting task as a regression problem (instead of segmentation and post-counting problem), by generating spatial density prediction maps. Later, several works presented a modified U-Net (Ronneberger et al., 2015) deep learning model for the segmentation of nuclei from bigger patches in nuclear IHC images (Zhang et al., 2020b) and the segmentation of cell membrane immunostaining in HER2 IHC images (Khameneh et al., 2019), avoiding the prerequisite of segmenting isolated cells. The results are more robust and computationally efficient than in previous works, but by solving the problem through a semantic segmentation, the algorithm has limitations in separating the grouped cells.

Emerging from the work reviewed, several obstacles to the development of advanced image processing techniques to address detailed marker quantification at IHC images can be identified. The main concern is related to the lack of labeled and publicly available data sets. Labeling a $1000 \times 1000$ pixel size region of interest in an $\times 40$ magnification image may involve the manual annotation of up to hundreds of cells. It is easy to understand the scarcity of databases containing this type of data, given the great effort required to generate them. Moreover, the patterns that can appear on the images can vary greatly depending on the type of cancer and its location, leading to a need for extensive training sets for the training of algorithms, further aggravating the aforementioned problem. In addition, the most recent described techniques treat the problem of cell segmentation as a problem of semantic segmentation. Semantic segmentation treats multiple objects of the same class as a single entity. However, instance segmentation treats multiple objects of the same class as distinct individual objects (or instances), which is ideal to separate immunopositive or immunonegative cells in cluttered areas.

As far as we know, this work presents the first method for precise and automated quantification of nuclear (ER, PR, Ki-67) and membrane (HER2) biomarkers using the same deep learning model structure that deals with instance segmentation of cells, where cells/nuclei of the same immunotype, although clustered, are unequivocally differentiated.

The major contributions of this paper include:

- Creation of training and test image datasets. We have developed a web-based platform, including a WSI viewer and annotation tool that allows pathology specialists to annotate IHC images, establishing in this way a method-ology to extract expert knowledge in the form of training data sets, to alleviate the tedious work of manual image annotation and to validate the development of new image processing methods.

- The design of an expert system for the accurate automatic quantification of digital breast cancer immunohistochemistry images through a computationally efficient and robust deep-learning based instance segmentation method capable of tackling the presence of clustered or overlapping cells as well as the presence of stromal cells and lymphocytes which are not subject to counting.

The reminder of this article is organized as follows. First, Section 2 introduces the description of the training and test data sets used in this work, as well as the annotation and decision support tools developed to create the above data sets. Then, the deep learning-based instance segmentation model used for the quantification of IHC images is described. Afterwards, Section 3 presents the experimental results of applying the instance segmentation method to nuclear and membrane IHC-stained images. Finally, Section 4 summarizes the limitations of the research, some concluding observations and future research directions.

## 2. Material and methods

### 2.1. Dataset description

The present study has been carried out on surgical specimens with histopathological diagnosis of breast carcinoma of different histological types. IHC biomarker quantification and scoring is restricted to:

- Nuclear markers ER, PR and Ki-67. ER expression is a favorable prognostic factor and strongly predictive of a response to hormone therapy. PR expression has traditionally been thought to signify functional ER and a positive response to endocrine therapy (Manni et al., 1980). Ki-67 assessment is the most commonly used method in clinical practice to determine proliferative activity in breast cancer. For the evaluation of ER, PR and Ki-67, the percentage of positively stained tumor cell nuclei is assessed, considering that the determination of the proportion of stained cells should be restricted to malignant cells, carefully avoiding tumor-infiltrating stroma and inflammatory cells.
- Membrane marker HER2. Along with hormone receptors, HER2 is the most important prognostic and predictive marker in breast cancer: HER2 amplification and overexpression is associated with adverse prognosis and is associated with high tumor grade, lymph node metastasis, high mitotic count and resistance to endocrine therapy (Hou et al., 2020). HER2 scoring ranges from 0 to 3+, with cases with a score of 0 or 1+ being classified as negative, 2+ cases as equivocal and cases with a score of 3+ as positive. Recommended guidelines for HER2 IHC scoring criteria for assessing HER2 status (Wolff et al., 2018), based on the intensity and integrity of cell membrane immunostaining, as well as the percentage of membrane-positive cells, are summarized in Table 1.

Table 1: Recommended HER2 scoring criteria for IHC assay of the invasive component of a breast cancer specimen.

| Score | Cell membrane staining pattern | Staining assessment |
|---|---|---|
| 0 | No staining is observed or membrane staining that is incomplete and is faint/barely perceptible and in ≤ 10% of tumor cells | Negative |
| 1+ | Incomplete membrane staining that is faint/ barely perceptible and in > 10% of tumor cells | Negative |
| 2+ | Weak to moderate complete membrane staining observed in > 10% of tumor cells | Equivocal |
| 3+ | Circumferential membrane staining that is complete, intense and in > 10% of tumor cells | Positive |

For this work, IHC techniques have been applied to 3-µ*m*-thick sections using the EnVision FLEX visualization Systems for Autostainer Link 48 (Agilent Technologies Inc.).

Tissues were counter-stained with hematoxylin and the chromogenic substrate used was diaminobenzidine (DAB), according to the supplier's instructions and the optimized protocol of the anatomic pathology unit. Samples were incubated with Ki-67 (Clone MIB-1), ER (Clone 1D5), PR (Clone PgR 636) and HER2 (HercepTest) antibodies.

IHC slides were then digitized using a PANNORAMIC 250 (3DHISTECH Ltd., Budapest, Hungary) bright-field scanner with a resolution of 0.25 µ*m*/pixel (equivalent to ×40 magnification). From a total of 23 IHC WSI images, pathologists and histo-technicians followed an agreed protocol and used a web-based WSI viewer to make annotations using freehand drawings in 350×350 pixel patches, by delineating tumor and non-tumor nuclei, as well as associating the appropriate category with them. The total number of patches annotated from nuclear and membrane IHC-stained images was 310 and 308, respectively.

Table 2: Breakdown of the number of annotated cells in training and test datasets.

(a) Cells extracted from HER2 IHC-stained images.

| | Train | Test | Total |
|---|---|---|---|
| No membrane staining | 2994 | 624 | 3618 |
| Barely perceptible and incomplete membrane staining | 340 | 160 | 500 |
| Weak to moderate and complete membrane staining | 804 | 189 | 993 |
| Intense and complete membrane staining | 1137 | 256 | 1393 |
| **All** | 5275 | 1229 | 6504 |

(b) Cells extracted from Ki-67, ER, and PR (nuclear) IHC-stained images.

| | | Train | Test | Total |
|---|---|---|---|---|
| Ki-67 | Immunopositive cells | 371 | 71 | 386 |
| | Immunonegative cells | 2421 | 619 | 3040 |
| ER | Immunopositive cells | 1088 | 281 | 1369 |
| | Immunonegative cells | 1280 | 327 | 1607 |
| PR | Immunopositive cells | 1024 | 268 | 1292 |
| | Immunonegative cells | 787 | 185 | 972 |
| **All** | | 6915 | 1751 | 8666 |

Table 2 shows the number of annotated cells in training and test datasets for both nuclear and membrane IHC-stained WSI images. Some examples of red-green-blue (RGB) image patches and their corresponding truth segmentation maps for the different cell classes categorized in nuclear and membrane IHC images are shown in Fig. 1 and Fig. 2, respectively. The ground-truth segmentation maps were defined as the reference maps outlined manually by the pathology experts.

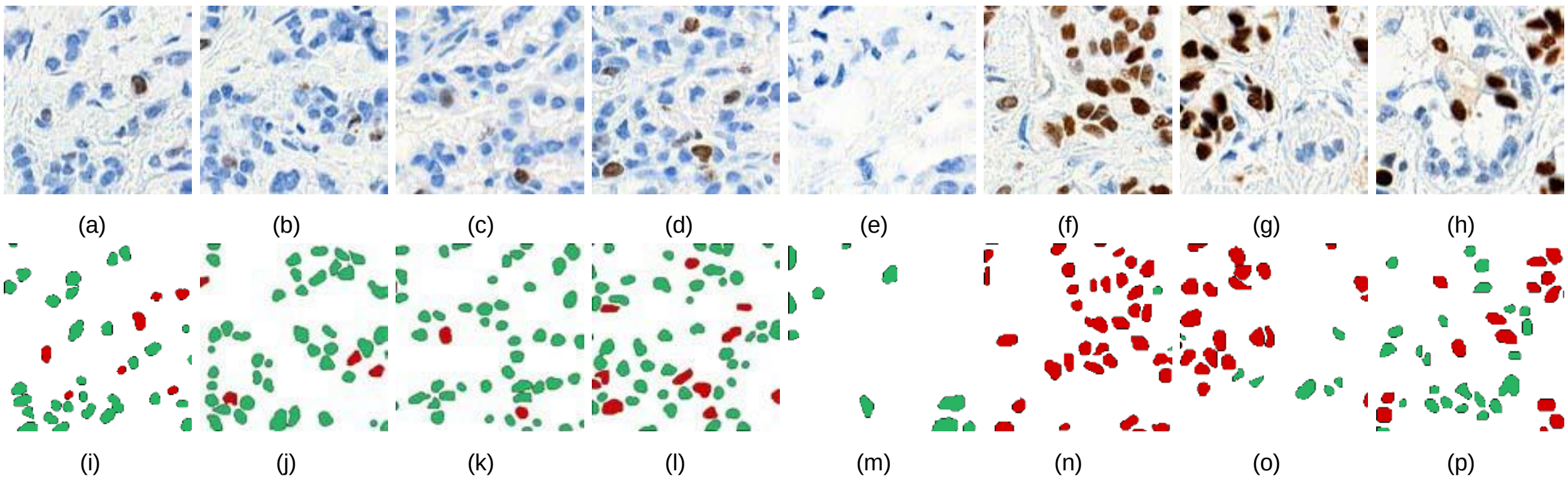

Figure 1: Samples of Ki-67 (a, b, c, d), ER (e, f) and PR (g, h) IHC images and their corresponding ground-truth segmentation maps (i, j, k, l, m, n, o, p) for the different types of nuclear staining: immunopositive cells and immunonegative cells.

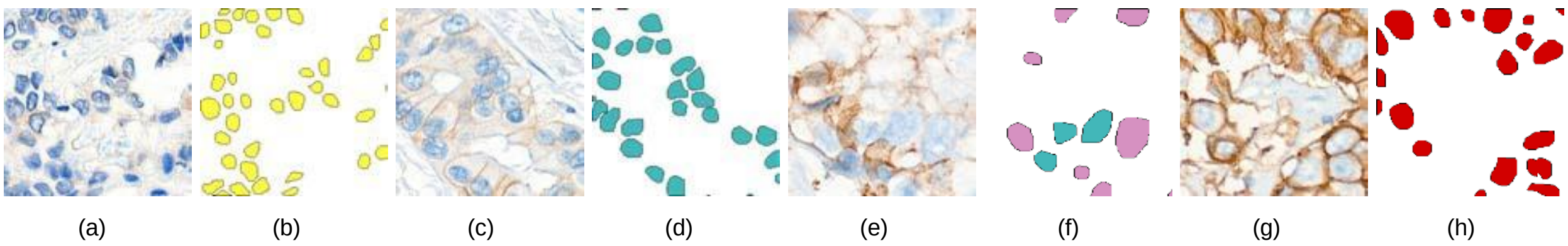

Figure 2: Samples of HER2 IHC images (a, c, e, g) and their corresponding ground-truth segmentation maps (b, d, f, h) for the different types of membrane staining: no membrane staining, barely perceptible and incomplete membrane staining, weak to moderate and complete membrane staining and intense and complete membrane staining.

## 2.2. Annotation and decision support tool

One of the most arduous tasks prior to training machine learning algorithms is the extraction of expert medical knowledge in the form of training and test datasets, in this case, a set of manually annotated IHC images. This requires the use of a software solution that minimizes pathologists' effort, is adapted to the WSI visualization problem, and has intuitive and easy-to-use annotation tools. For a previous work (Priego-Torres et al., 2020), we developed a web-based, vendor-neutral digital slide viewer with annotation tools adapted to the segmentation problem of whole-slide H&E stained breast histopathology images. Given the successful performance of this previous work, we adapted this viewer to integrate IHC images, which have the particularity of including thousands of small annotations around cells. Again, this web viewer allowed us to establish a procedure for dataset generation, enabling a rapid transition from annotation to the development of new image processing techniques and the validation and visualization of the results. Table 3 details the specifications of the technologies involved in the development of the web-based annotation and decision support tool.

Table 3: Specifications of the technologies involved in the development of the annotation and decision support tool.

| | |
|---|---|
| Backend web framework | Python Flask |
| Patient information and WSI hosting | Images locally hosted<br>Data managed with a SQLite database locally |
| WSI Image accessing | OpenSlide image library |
| WSI web visualization | OpenSeadragon library (JavaScript) |
| Annotations | Vector objects drawn with Paper.js (JavaScript) |
| Annotation storage format | JSON |

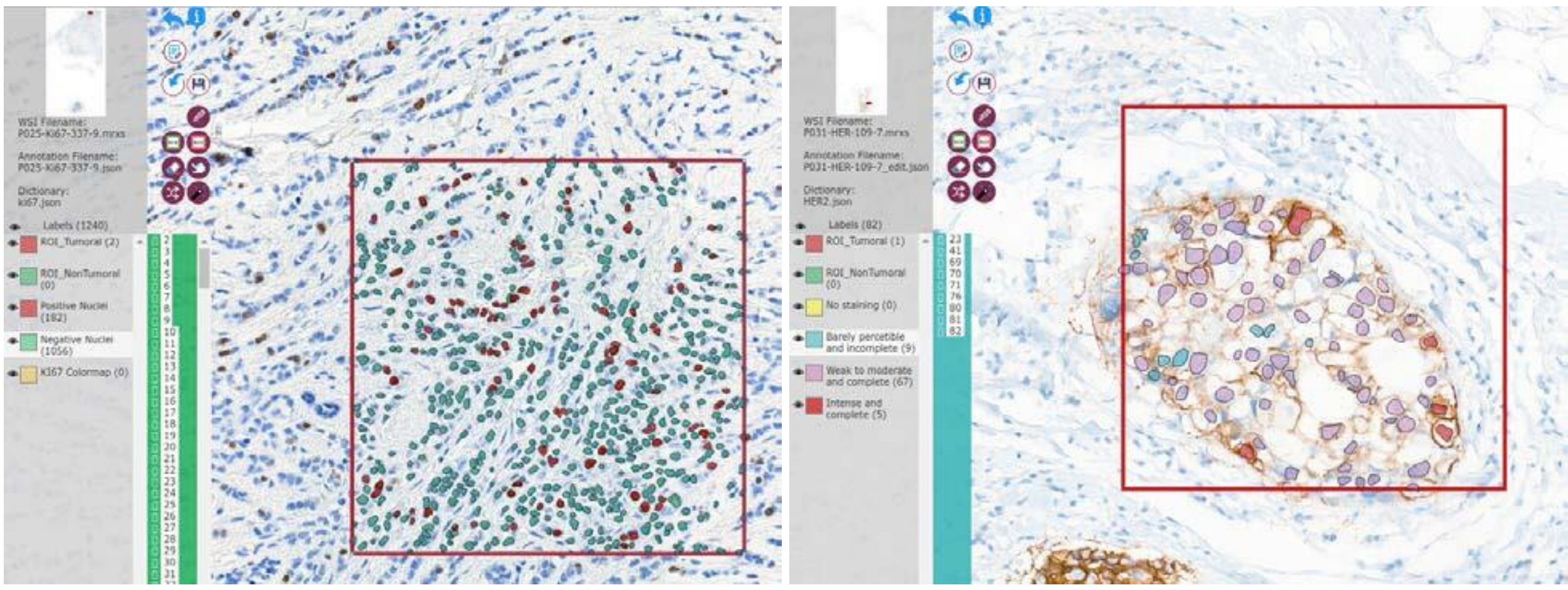

(a) (b)

Figure 3: Screenshots of the annotation and decision support tool.

The WSI visualization, annotation and decision support tool has been a pivotal point for the generation of training sets that have allowed the tuning of the nuclear and membrane biomarker quantification models. Inspired by work in the literature using cascade learning or iterative verification (Tajbakhsh et al., 2020; Xue et al., 2020; Su et al., 2015), the generation of the set of labeled images has been obtained progressively and following a pseudo-labelling approach (Lee et al., 2013): in the first phase, a minimal set of images were manually labeled by expert pathologists. This set was created freehand and at the maximum magnification allowed by the WSI viewer (0.25 $\mu m$/pixel, equivalent to $\times$40 magnification), in order to minimise inter-observer error interpretation. It is important to point out that although pathologists could zoom in and out on the images for inspection at different levels of resolution, the freehand annotation tool was always used at the highest level of magnification. To this end, it was agreed that the pathologists in charge of annotation would follow the same annotation protocol and use the same hardware equipment. From this set, a first version of the deep learning-based instance segmentation models were trained. Then, the resulting trained models were integrated into the annotation tool as a decision support add-on. In a second phase of image labeling, instead of annotating new datasets from scratch, we started from a pre-segmentation of instances obtained by the pre-trained models, so that the task of the expert user was limited to correcting those invalid cells segmentations. Again, corrections after pre-segmentation were made by pathologists freehand at μm/pixel. This procedure was repeated several times to obtain the final training and test data set used in this work to validate the proposal for accurate biomarker quantification.

Fig. 3 shows screenshots of the tool, with details on the annotation utilities and the decision support button to automatically segment cells in both nuclear and membrane stained images. The video included as supplementary material in this publication illustrates the use of the developed tool.

### *2.3. Deep-neural-network based instance segmentation architecture*

This study aims to segment and categorize malignant cells in IHC images for subsequent quantification and scoring. As previously mentioned, a particularity of this type of stained pathology images is that cells that need to be counted separately may appear clustered. With this in consideration, we decided to approach this image processing task by applying instance-based segmentation techniques instead of the more commonly used semantic segmentation ones. While semantic segmentation consists in detecting objects in an image and grouping them according to classes or categories, instance segmentation takes semantic segmentation one step further, performing the detection of instances within defined categories, so that they are uniquely labeled.

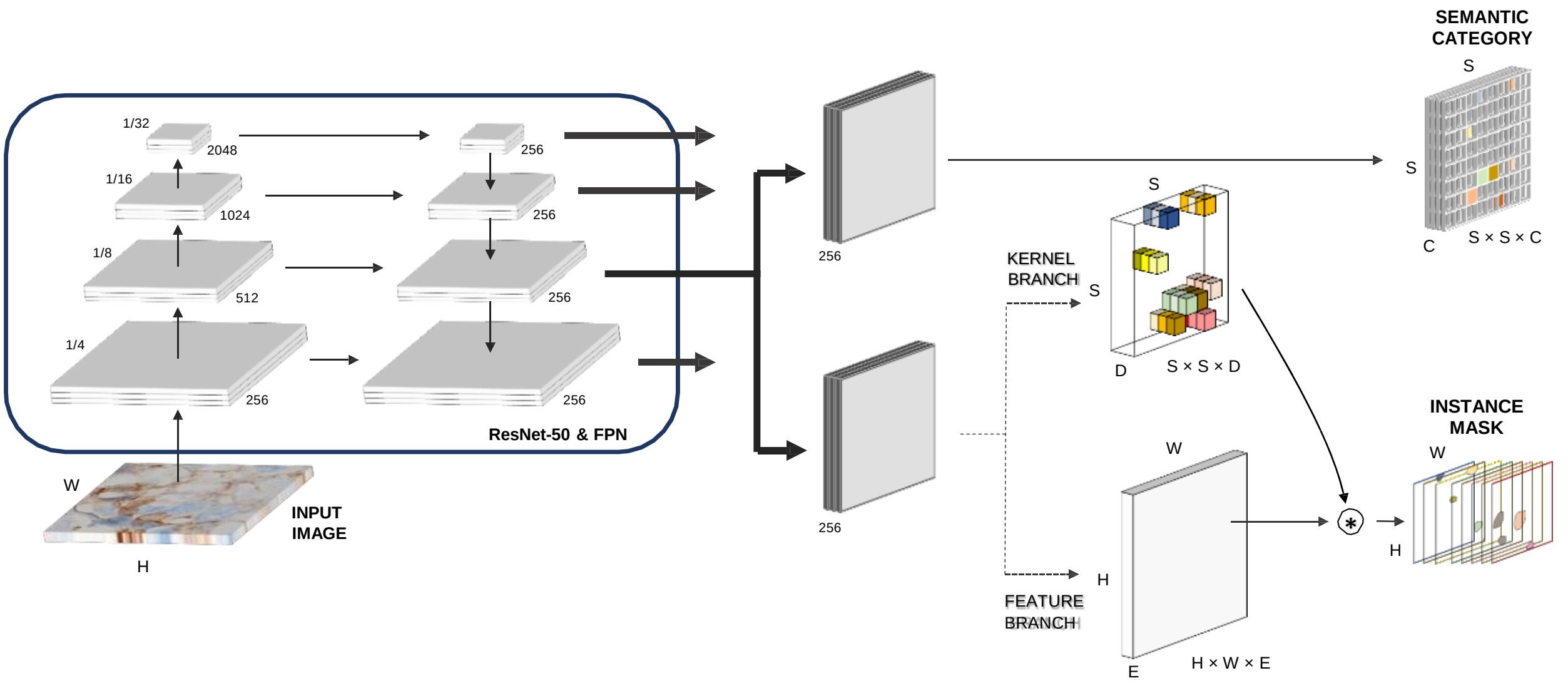

Figure 4: SOLOv2 architecture.

Instance segmentation is gaining importance and is becoming a complex and challenging field of computer vision research. With the advent of deep learning (DL), more specifically, convolutional neural networks (CNNs), multiple instance segmentation methods have been proposed in the last four years (He et al., 2017; Wang et al., 2020c; Zhang et al., 2020a; Bolya et al., 2019; Chen et al., 2019b, 2020). All these algorithms face the major challenges of instance segmentation, which can be summarized as delineating objects with sufficient detail; handling occlusions, since they often lead to loss of information of object instances, which in IHC images occurs when cells appear cluttered; detecting small objects, as well as objects at various scales and geometric transformations; and finally, that algorithms are able to process images in a time-efficient manner.

After reviewing the state-of-the-art of instance segmentation algorithms, in this work we selected the SOLO method (Wang et al., 2020a,b), specifically its second version SOLOv2, which is presented as more precise and faster, to solve the problem of segmentation and categorization of cells for subsequent quantification in IHC images. SOLOv2 was chosen because, on the one hand, it presents excellent performance values when standard image sets are applied and because of the low computational cost which translates into low processing times. On the other hand, it is perfectly adapted to our cell segmentation problem, characterized by having a large number of instances in the same image and the presence of occlusions. In addition, to ensure the suitability and high performance of SOLOv2 in the calculation of scoring in immunohistochemistry images in terms of the selected evaluation metrics, we compared the results obtained using SOLOv2 with a well-established benchmark instance segmentation algorithm, Mask-RCNN (He et al., 2017), of which multiple variants have emerged in recent years.

The underlying philosophy that defines SOLOv2 is very simple and effective: any two instances in an image can be effectively distinguished by their size and their center locations. In this way, the instance segmentation problem is reformulated as two simultaneous category-aware prediction problems.

In a first processing step, SOLOv2 relies on a convolutional backbone network, specifically a ResNet-50 model (He et al., 2016), followed by a Feature Pyramid Network (FPN) (Lin et al., 2017), a pyramidal hierarchy of deep convolutional networks to build pyramids of feature maps with different sizes and a fixed number of channels at each level. Next, each of the feature maps is conceptually divided into $S \times S$ grids, where $S$ is different for each FPN level, considering that each grid cell is a potential target instance: if the center of an object falls into a grid cell, that grid cell is responsible for predicting the semantic class as well as the instance masks. To this end, two distinct sub-networks are attached to the FPN maps, one dedicated to semantically categorize each potential object in the image and the other to extract its instance mask. Fig. 4 illustrates the model architecture. Full implementation details of the deep learning-based architecture SOLOv2 can be found in (Wang et al., 2020a,b).

Mask-RCNN, the reference model used to contrast results, follows a top-down approach of segmentation-by-detection: in a first stage, it generates proposals of regions where objects or instances can be found based on features extracted from a backbone network, in this case, the ResNet-50 architecture. In a second stage, it predicts the class of the object, refines the bounding box and generates a mask of the instance based on the first stage proposal. Comprehensive details on the Mask-RCNN framework can be accessed at (He et al., 2017).

### 2.4. Transfer learning and data augmentation

One of the major problems that arise in the learning process of DL algorithms is that of adjusting a large number of learnable parameters (weights and biases) when the training dataset is not large enough. Generating massive hand-labeled training sets is usually expensive and, in most cases, requires expert knowledge as in the field of digital pathology.

This problem is addressed through transfer learning and data augmentation techniques. Using inductive transfer learning, a model is pre-trained using large labeled datasets from an unrelated problem and then adapted to the problem at hand with minor retraining, thus avoiding costly data labeling efforts. Particularly, in this work, all parameters of the ResNet-50 backbone network were previously pre-trained using as dataset the images provided for the COCO (Common Objects in Context) 2015 challenge (He et al., 2016). In the training, no parameter of any stage of the ResNet-50 backbone network was frozen. By the second technique, data augmentation, new versions of existing training images are created by applying image transformations. For this purpose, resizing at 6 different fixed scales and flipping with a probability of 0.5 has been considered.

### 2.5. Validation and evaluation metrics

To evaluate the performance of the instance segmentation algorithm, this study used several success rate metrics commonly applied in recent instance segmentation works to compare the consistency between automated quantification and ground truth.

In instance segmentation, the evaluation is not trivial, as there are two distinct tasks to assess: determining the presence of an object of a given type (classification) and accurately segmenting it in the image. In addition, the data set to be evaluated usually has more than one class, and the distribution of instances across classes is often not uniform. Due to these characteristics, one of the most commonly used metric for multi-class classification is the mean average precision ($mAP$) at Intersection Over Union ($IoU$).

To understand this measure, it is necessary to introduce some concepts. In a classification context, precision measures the ratio of true object detections to the total number of objects that the classifier predicted, while recall measures the ratio of true object detections to the total number of objects in the dataset. Thus, precision-recall is a useful measure of prediction success even when the classes are unbalanced, since precision gives a measure of the relevance of the results and recall measures how many truly relevant results are predicted. On the other hand, intersection over union ($IoU$) is the area of overlap between the predicted segmentation and the ground truth divided by the area of overlap between the predicted segmentation and the ground truth, where $IoU = 0$ means that there is no overlap and $IoU$ = 1 means that the segmentation overlaps perfectly. Formally, these metrics are expressed as:

$$Precision = \frac{T_p}{T_p + F_p}, \quad Recall = \frac{T_p}{T_p + F_n}, \quad IoU = \frac{Area\ of\ Overlap}{Area\ of\ Union} \qquad (1)$$

where $T_p$, $F_p$ and $F_n$ denote the number of true positives, false positives and false negatives detected in the image, respectively.

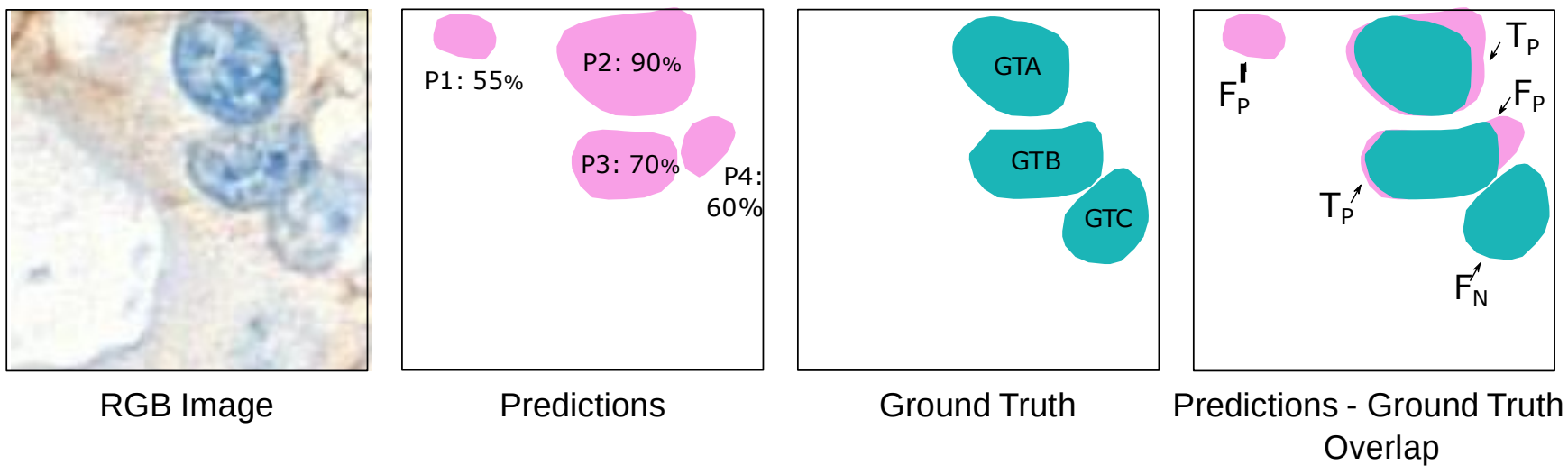

Figure 5: Instance segmentation example: the image contains 3 ground truths (GTA, GTB and GTC) and 4 regions are predicted (P1, P2, P3 and P4) for a particular class. The percentages associated to the prediction labels denote the confidence score provided by the instance segmentation model.

Now, let's consider the instance segmentation result depicted in Fig 5. For this example, we will set a $IoU_{th}$ threshold of 0.5 and a confidence score threshold ($\tau_{th}$) of 50% (provided by the instance segmentation model) for counting the prediction as $T_P$, with $F_N$ otherwise. Now, to calculate the value of $P_T$, $P_F$ and $N_F$ for a specific class, the following considerations are taken into account: when several segmentations detect the same object and their $IoU > IoU_{th}$, the one with the maximum confidence score is considered as $T_P$, while the rest are regarded as $F_P$; if an object is present and the $IoU$ between the predicted segmentation and the grouth truth object is less than $IoU_{th}$, the prediction is considered $F_N$; and if an object does not exist in the image, but the instance segmentation algorithm detects one, the prediction is considered $F_P$. Then, recall and precision are computed by applying the equations shown in (1), where predictions are sorted by their confidence levels and $T_P$ and $F_P$ values are accumulated. Considering the above, Table 4 summarises numerically the precision and recall metrics of the segmentation results pictured in Fig. 5. These precision and recall values can be plotted to obtain a PR (precision-recall) curve. To obtain the value of the average precision of a class in the data sets, the area under this PR curve has to be calculated using the Riemann integral method. Finally, after having one $AP$ per class, the $mAP$ is calculated as the average $AP$ of all existing categories in the dataset.

Table 4: Summary of the prediction and recall values of the example segmentation result shown in Fig. 5 considering $IoU_{th}$ = 0.50.

| Detection | Confidence | $IoU$ | Ground Truth | $T_P$ | $F_P$ | Precision | Recall |
|---|---|---|---|---|---|---|---|
| P2 | 90% | >0.5 | GTA | 1 | 0 | 1 | 0.33 |
| P3 | 70% | >0.5 | GTB | 2 | 0 | 1 | 0.33 |
| P4 | 60% | <0.5 | GTB | 2 | 1 | 0.67 | 0.67 |
| P1 | 55% | <0.5 | - | 2 | 2 | 0.5 | 0.67 |

In the previous example, a prediction with $IoU > 0.5$ has been considered to be a $T_P$. This results in a value of $mAP$ which is denoted as $mAP^{IoU=.50}$. However, the use of a fixed $IoU$ threshold may introduce a bias in the evaluation metric. For example, two predictions with different overlap, e.g. $IoU = 0.55$ and $IoU = 0.9$, will be weighted equally. This can be solved, instead of using a fixed $IoU_{th}$ value, by considering a range of $IoU$ thresholds, such that a $mAP$ is calculated for each $IoU_{th}$ and then averaged to obtain the final $mAP$. As a range of threshold values, the most commonly used (Lin et al., 2014) includes 10 thresholds between 0.5 and 0.95, and the $mAP$ value is then denoted as $mAP^{IoU=[0.50:0.05:0.95]}$. In this work we used $mAP^{IoU=0.50}$, $mAP^{IoU=0.75}$ and $mAP^{IoU=[0.50:0.05:0.95]}$ metrics to validate cell counts in the biomarker quantification process.

Further details and rigorous mathematical definitions of these instance segmentation evaluation metrics can be found in (Padilla et al., 2021).

Having introduced these metrics, it is important to note that, in addition to having a higher level of complexity than those used in semantic segmentation, which are usually based on *IoU* or pixel accuracy, the *mAP* values are often relatively low in comparison. Figure 6 illustrates an example of this. If we assume that the prediction in this figure addresses the instance segmentation problem, and we consider a $IoU_{th}$ = 0.50, the prediction results will be considered as false negatives. However, by applying metrics specific to semantic segmentation, the result for pixel accuracy and *IoU* will be close to 85% and 0.85, respectively, which implies a significant quantitative difference.

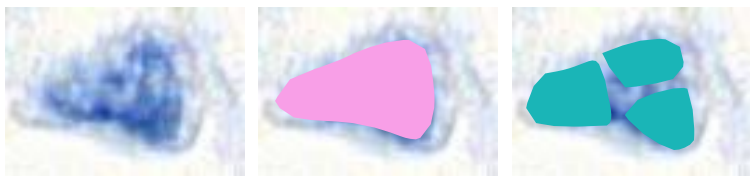

Figure 6: Segmentation example: ground truth and prediction.

### 2.6. System configuration

Training and tests of the quantification algorithm applied to Ki-67, ER, PR and HER2 stained breast histopathology images was accelerated by using an NVIDIA DGX station, made up of four NVIDIA® Tesla® V100 Tensor Core GPUs, integrated with a fully-connected four-way NVLink™ architecture.

#### 2.6.1. SOLOv2-based segmentation model configuration

The instance segmentation approach was based on the publicly available model SOLOv2 [1], implemented on Pytorch framework, that relies on ResNet-50 externally pre-trained on COCO 2015 dataset as backbone network. For the training of SOLOv2, a stochastic gradient descent (SGD) method was used with an initial learning rate, weight decay and momentum of 0.01, 0.0001 and 0.9, respectively. The number epochs used for training was 120. Each experimental learning process took 1 hour and 40 minutes approximately and the inference time for each image in the test dataset is about 0.15 seconds.

The loss function used in the training process is defined as $L = L_{cate} + L_{mask}$ where $L_{cat}$ is the conventional Focal Loss for semantic category classification and $L_{mask}$ is the Dice Loss for mask prediction (Wang et al., 2020a). Fig. 7 depicts the learning curves in the training process, showing the losses $L_{mask}$, $L_{cate}$ and the sum of both, $L$. The loss dropped rapidly below 0.6 for the nuclei-aware segmentation model after approximately 34 epochs (around 5.000 steps) and below 0.7 after about 40 epochs (around 4.000 steps), for the membrane-aware segmentation model, stabilizing after that point.

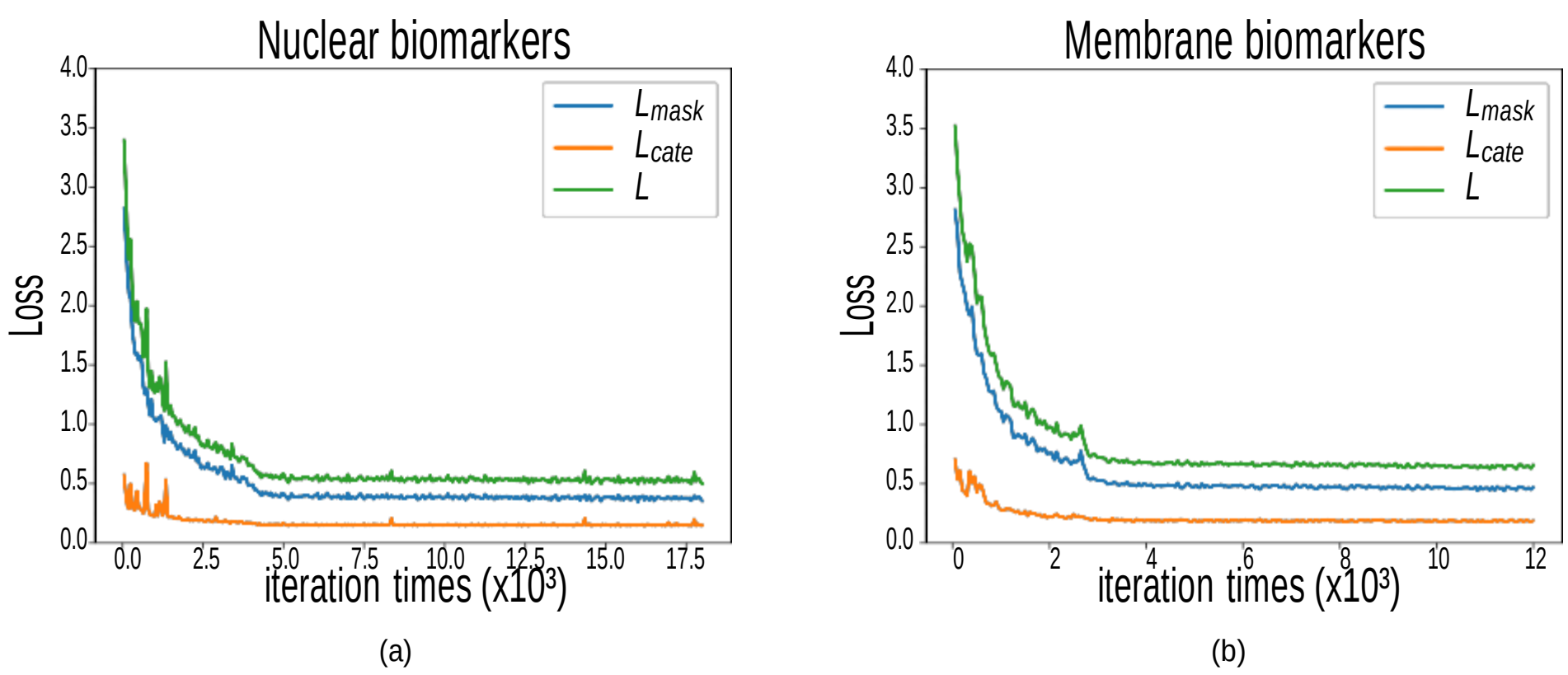

Figure 7: Training curve for both nuclei- and membrane-aware segmentation models based on SOLOv2.

[1] https://github.com/WXinlong/SOLO/blob/master/configs/solov2/solov2_r50_fpn_8gpu_3x.py

### 2.6.2. Mask-RCNN-based segmentation model configuration

The Mask-RCNN-based segmentation model was extracted from the publicly available model Mask-RCNN [2], implemented on TensorFlow, and relying on ResNet-50 externally pre-trained on COCO 2015 dataset as backbone network. The nuclei- and membrane-aware Mask-RCNN-based models were trained for 120 epochs and took 3 hours and 30 minutes approximately. A stochastic gradient descent (SGD) method was also adopted, with a learning rate of 0.003 and momentum of 0.9. The inference time for each image in the test dataset is approximately 0.3 seconds.

In the learning process a multi-task loss is used, defined as $L = L_{cls} + L_{box} + L_{mask}$, where $L_{cls}$ is the classification loss, $L_{box}$ is the regression loss of bounding boxes (Zhang et al., 2016) and $L_{mask}$ denotes the average binary cross-entropy loss, only including $k$-th mask if the region is associated with the category $k$.

## 3. Results and discussion

### 3.1. Instance segmentation and IHC quantification assessment

Throughout this results section, we will refer to the nuclei-aware segmentation model as the SOLOv2-based instance segmentation algorithm trained to segment nuclei in nuclear-stained images (Ki-67, ER and PR IHC images), whereas the membrane-aware segmentation model will denote the one trained to detect nuclei in membrane-stained images (HER2 IHC images).

A first experiment was conducted to optimize the confidence score threshold, $\tau_{th}$, which determines whether an instance segmented by SOLOv2-based models should be considered as a tumoral nucleus. For this purpose, a parameter sweep was performed on parameter $\tau_{th}$. Then, the *mAP* values of the instance segmentation maps resulting from applying the nuclei- and membrane-aware segmentation models on the test dataset for each of the $\tau_{th}$ values were calculated. The outcomes of this experiment are plotted in Fig. 8, where it can be noted that the best result for the $mAP^{IoU=0.5}$ metric is obtained when $\tau_{th}$ is around 0.3 (30%) for both nuclei- and membrane-aware segmentation models. Therefore, in subsequent experiments, $\tau_{th}$ was set to this value.

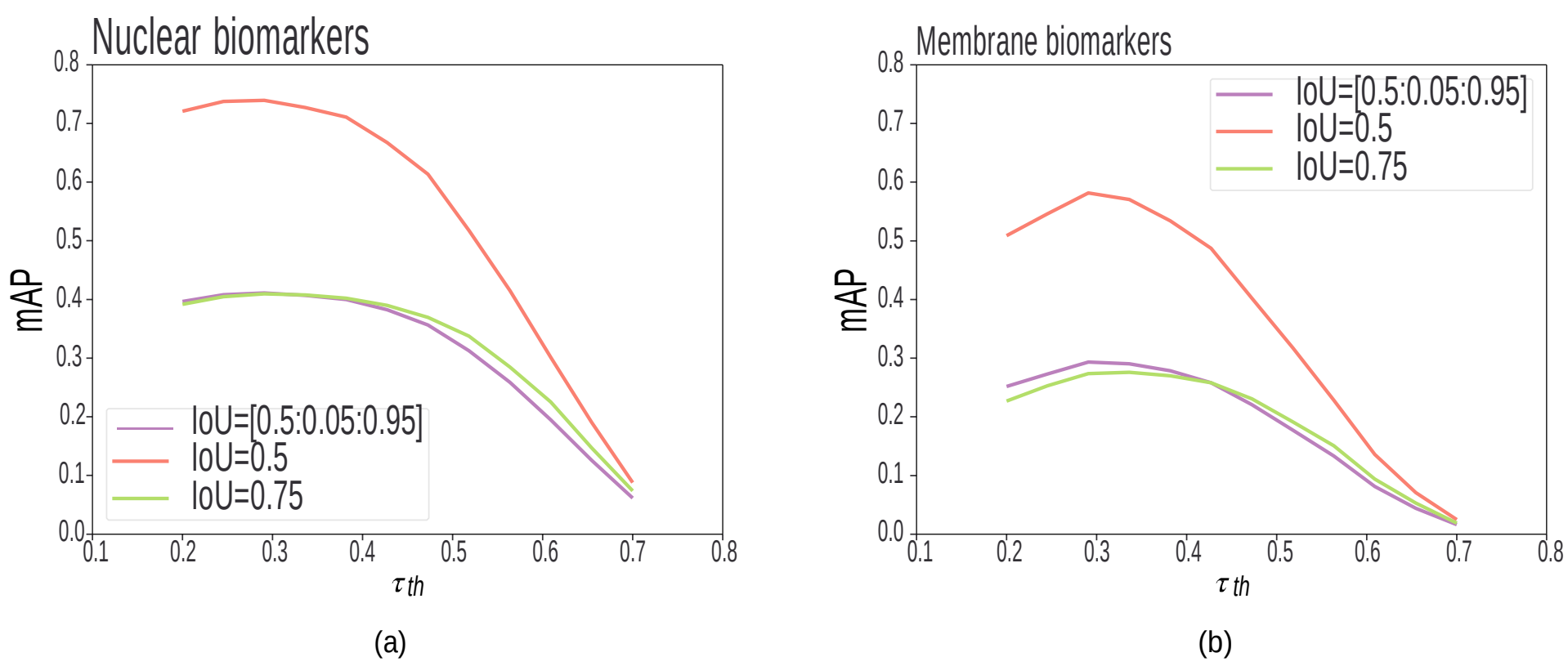

Figure 8: *mAP* versus $\tau_{th}$ for both nuclei- and membrane-aware segmentation models applied to the test image sets.

In the next step, the instance segmentation metrics considered in this study were calculated for the validation of the nuclei- and the membrane-aware segmentation models over the test sets, in the former case also differentiating the performance results according to the types of nuclear staining. A detailed analysis of the results is provided in Tables 5 and 6.

As for the segmentation results for the nuclei-aware segmentation model, a $mAP^{IoU=.50}$ value of 0.67, 0.77 and 0.72 is achieved for the test Ki-67, ER and PR IHC-stained image dataset, respectively. The results in Table 5 also indicate that immunonegative cells were, in general, slightly more reliably segmented than immunopositive cells.

Tile captures in Fig. 9 demonstrate how instance segmentation maps of tumor cell instances closely follow the expert pathologists' segmentation criteria for nuclear IHC imaging. It is noteworthy how neatly the model segments images Fig. 9a-d, which exhibit patches of different Ki-67 IHC-stained images, where high heterogeneity in terms of intensity, hue and texture of the tumour nuclei can be observed. Another highlight is the model's ability to segment instances when they appear grouped together, as is the case in Fig. 9g, extracted from an PR IHC-stained image, where immunopositive nuclei lie cluttered. In addition, the model avoids the segmentation of stromal cells and lymphocytes, which are not subject to counting in the biomarker quantification process, as can be seen in Fig. 9e and 9g.

[2] https://github.com/matterport/Mask_RCNN

Table 5: Assessment of the nuclei-aware segmentation model applied to Ki-67, ER and PR IHC-stained images.

| **Biomarker** | $mAP^{IoU=.50}$ | $mAP^{IoU=.75}$ | $mAP^{IoU=[.50:.05:.95]}$ |
|---|---|---|---|
| Ki-67 - All tumor cells | 0.67 | 0.35 | 0.36 |
| Ki-67 - Immunopositive cells | 0.59 | 0.22 | 0.27 |
| Ki-67 - Immunonegative cells | 0.74 | 0.48 | 0.43 |
| ER - All tumor cells | 0.77 | 0.46 | 0.45 |
| ER - Immunopositive cells | 0.75 | 0.36 | 0.39 |
| ER - Immunonegative cells | 0.78 | 0.56 | 0.50 |
| PR - All tumor cells | 0.72 | 0.40 | 0.41 |
| PR - Immunopositive cells | 0.72 | 0.30 | 0.35 |
| PR - Immunonegative cells | 0.72 | 0.49 | 0.46 |

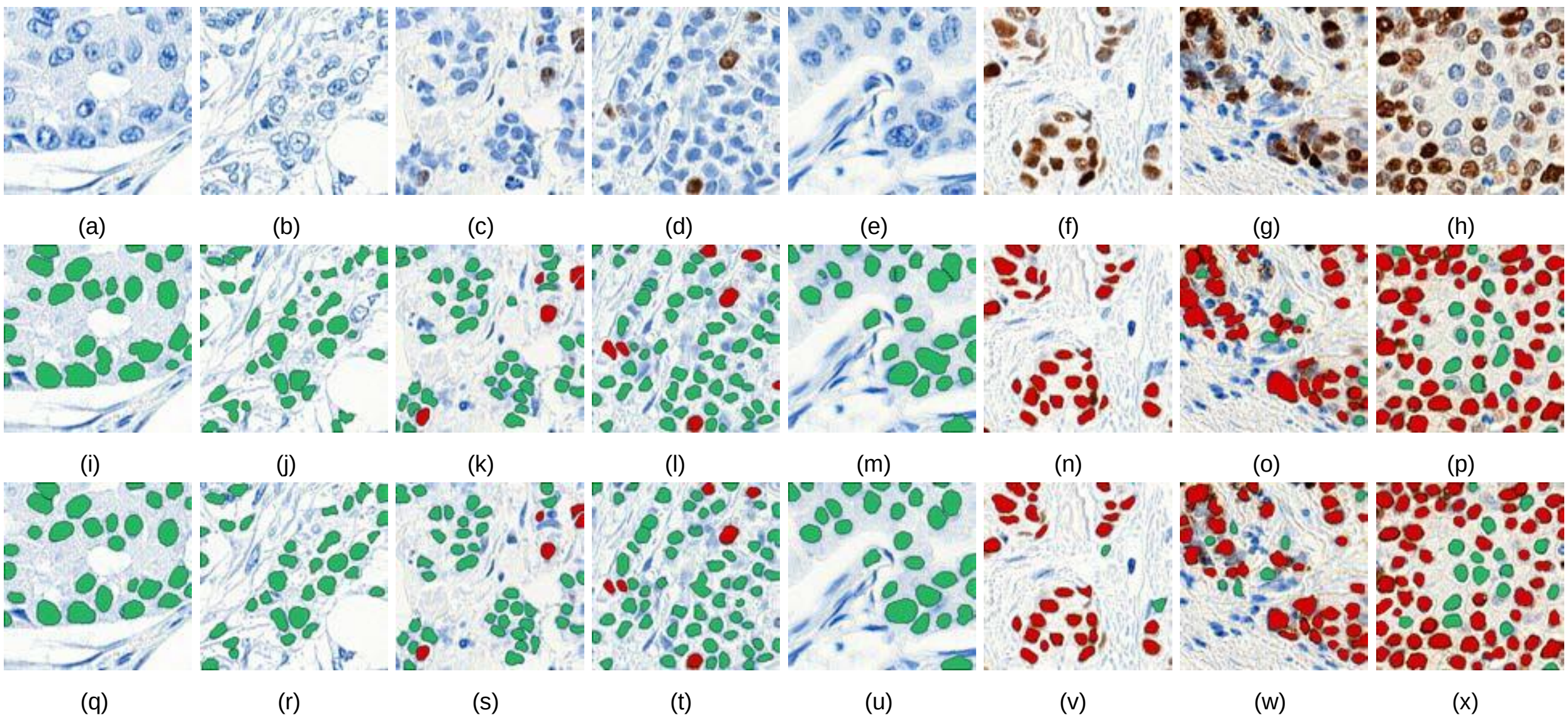

Figure 9: Samples of Ki-67 (a, b, c, d), ER (e, f) and PR (g, h) IHC-stained images, their corresponding ground-truth (i, j, k, l, m, n, o, p) and output instance segmentation maps (q, r, s, t, u, v, w, x) for the different types of nuclear staining: immunopositive cells and immunonegative cells.

Fig. 10 shows series of precision-recall curves related to the segmentation of instances for both immunopositive and immunonegative nuclei classes, considering separately each type of nuclear staining. For each subplot, $IoU = 0.5$ and $IoU = 0.75$ were used to extract these curves, as well as *Oth*, which represents instances that are confused with other categories, in this case, immunopositive nuclei with immunonegative and vice versa.

For the Ki-67 test set, the distance between the $IoU = 0.75$ and $IoU = 0.50$ curve when assessing the immunopositive nuclei category (Fig. 10.a) is evident. This denotes that nuclei are effectively detected but the overlap between the ground truth and the segmentation provided by our model differs, leading to the detection of nuclei with larger or smaller size or shape. On the other hand, the *Oth* and $IoU = 0.50$ curves closely match, indicating that there are few immunopositive nuclei that have been classified as negative in the test dataset. As for graph (Fig. 10.d), which evaluates the precision and recall for immunonegative nuclei in the test set of Ki-67 IHC-stained images, a better segmentation performance is observed, although in this case a misclassification of negative nuclei as positive does appear.

A similar behaviour is seen in the plots related to the assessment of precision-recall for the immunopositive (Fig. 10.b) and immunonegative (Fig. 10.e) nuclei categories, as well as for the immunopositive (Fig. 10.c) nuclei category, in ER and PR IHC-stained test images, respectively. Finally, graph Fig. 10.f illustrates how immunonegative nuclei in PR-stained images are appropriately segmented and that even misclassification with positive nuclei is rather low.

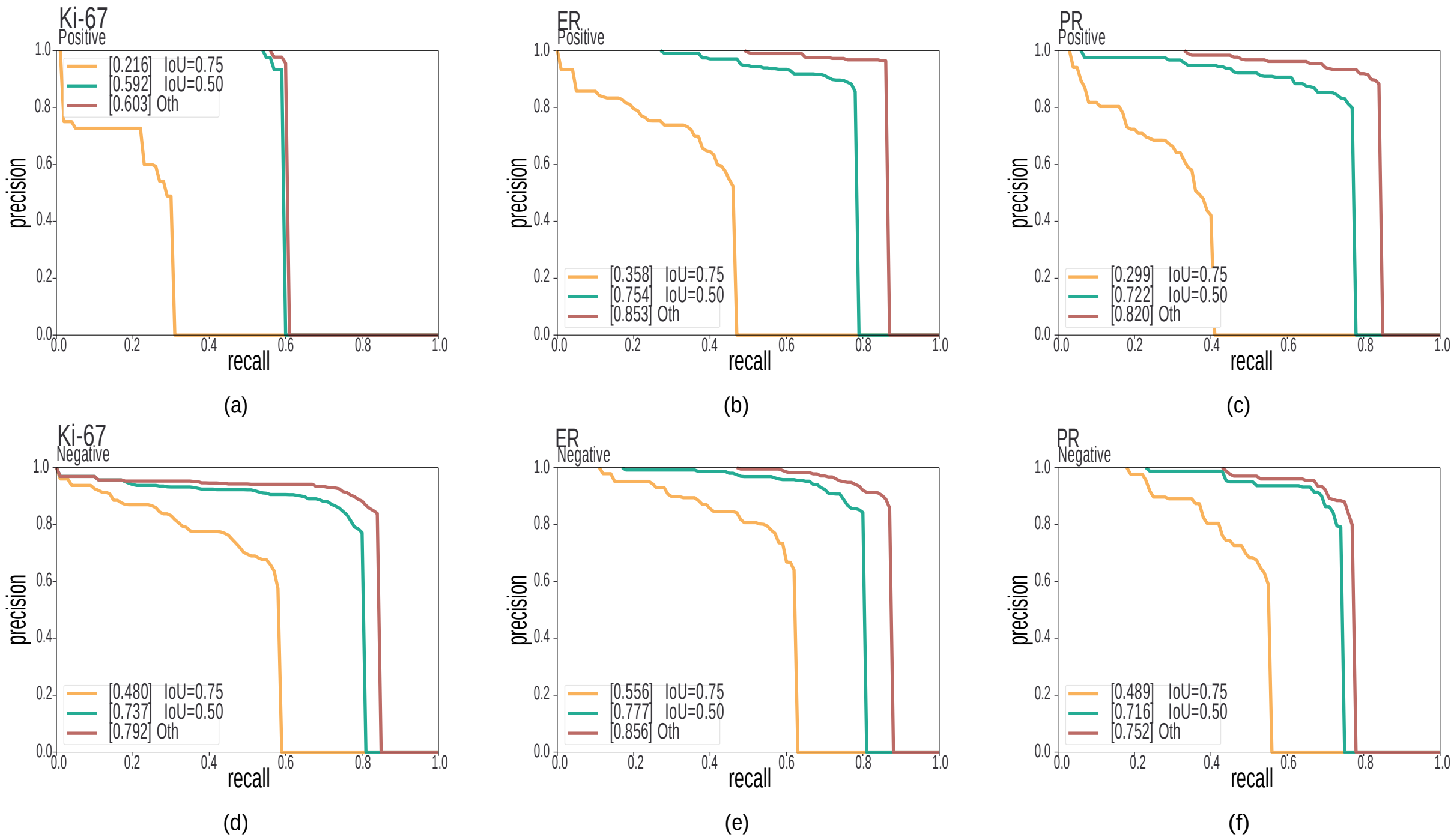

Figure 10: Precision-recall curves ($IoU$ = 0.5, $IoU$ = 0.75 and $Oth$) for immunopositive (a, b, c) and immunonegative (d, e, f) nuclei categories considering test sets extracted from Ki-67 (a, d), ER (b, e) and PR (c, f) IHC-stained images, respectively.

Moving on to the evaluation of the membrane-aware segmentation model applied to HER2-stained images, Table 6 reports the AP values for the different categories of nuclei according to the staining and appearance of their membranes. It is observed that the $mAP^{IoU=.50}$ in this test set is 0.56, with the maximum value of $AP^{IoU=.50}$ for the category of “circumferential, intense and complete membrane staining” being 0.74, and the minimum value for the category of “barely perceptible and incomplete membrane staining” dropping to 0.22. These results are consistent with the output segmentation maps depicted in Fig. 11. Again, the tiles selected as representative samples show variability in terms of shape, size and texture of nuclei and in this case also of membrane, as well as the presence of other biological structures. A visual evaluation of the segmentation maps in comparison to their ground truth shows that the membrane-aware segmentation model works satisfactorily, except in the categorisation of nuclei with barely perceptible and incomplete membrane, which tends to be confused with the “no membrane staining” class, as shown in shown in Fig. 11.q and 11.s.

Table 6: Assessment of the membrane-aware segmentation model applied to HER2 IHC-stained images.

| **Biomarker** | $mAP^{IoU=.50}$ | $mAP^{IoU=.75}$ | $mAP^{IoU=[.50:.05:.95]}$ |
|---|---|---|---|
| HER2 - All tumor cells | 0.56 | 0.26 | 0.29 |
| HER2 - 0: No membrane staining | 0.66 | 0.27 | 0.32 |
| HER2 - 1+: Barely perceptible and incomplete membrane staining | 0.22 | 0.08 | 0.10 |
| HER2 - 2+: Weak to moderate and complete membrane staining | 0.63 | 0.36 | 0.35 |
| HER2 - 3+: Circumferential, intense and complete membrane staining | 0.74 | 0.34 | 0.37 |

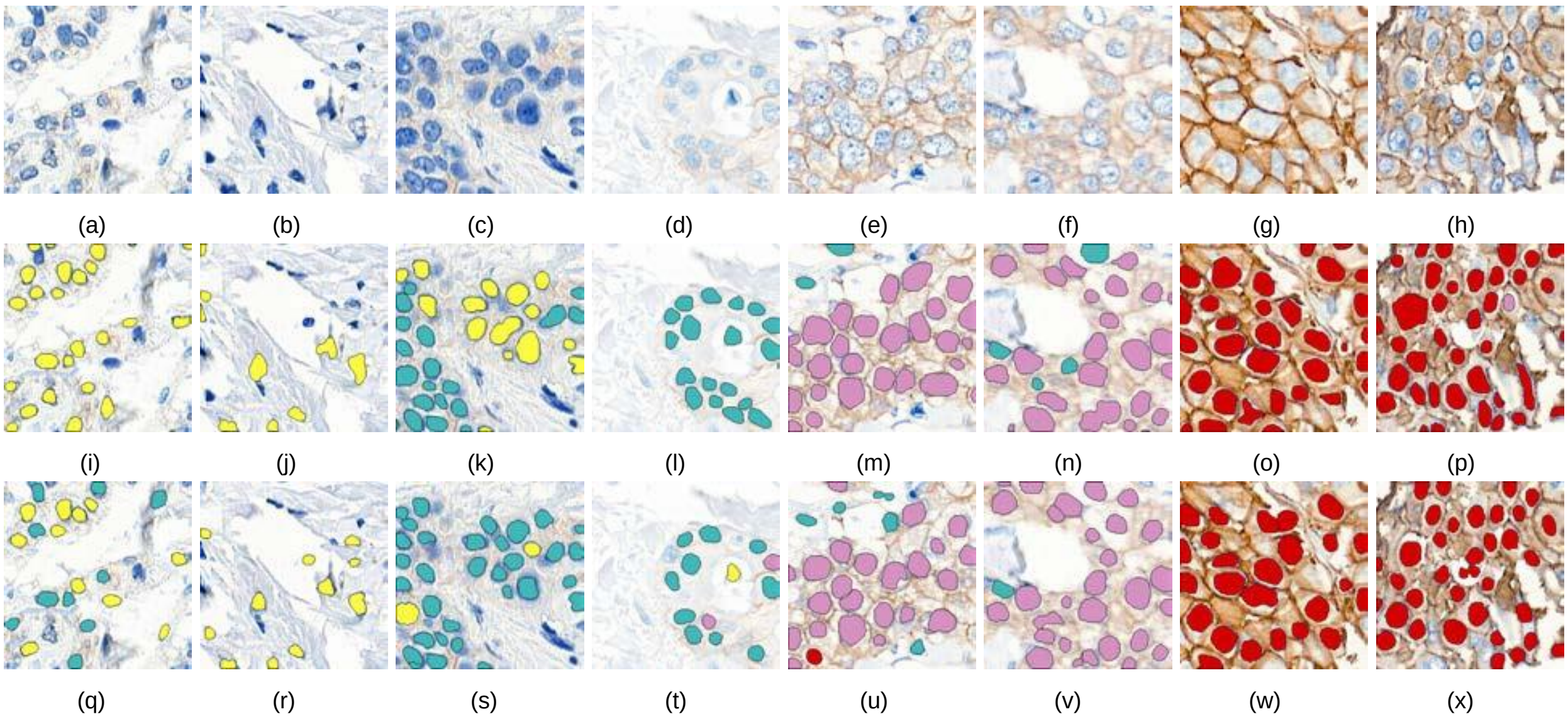

Figure 11: Samples of HER2 IHC images (a, b, c, d, e, f, g, h), their corresponding ground-truth (i, j, k, l, m, n, o, p) and output instance segmentation maps (q, r, s, t, u, v, w, x) for the different types of membrane staining: no membrane staining, barely perceptible and incomplete membrane staining, weak to moderate and complete membrane staining and intense and complete membrane staining.

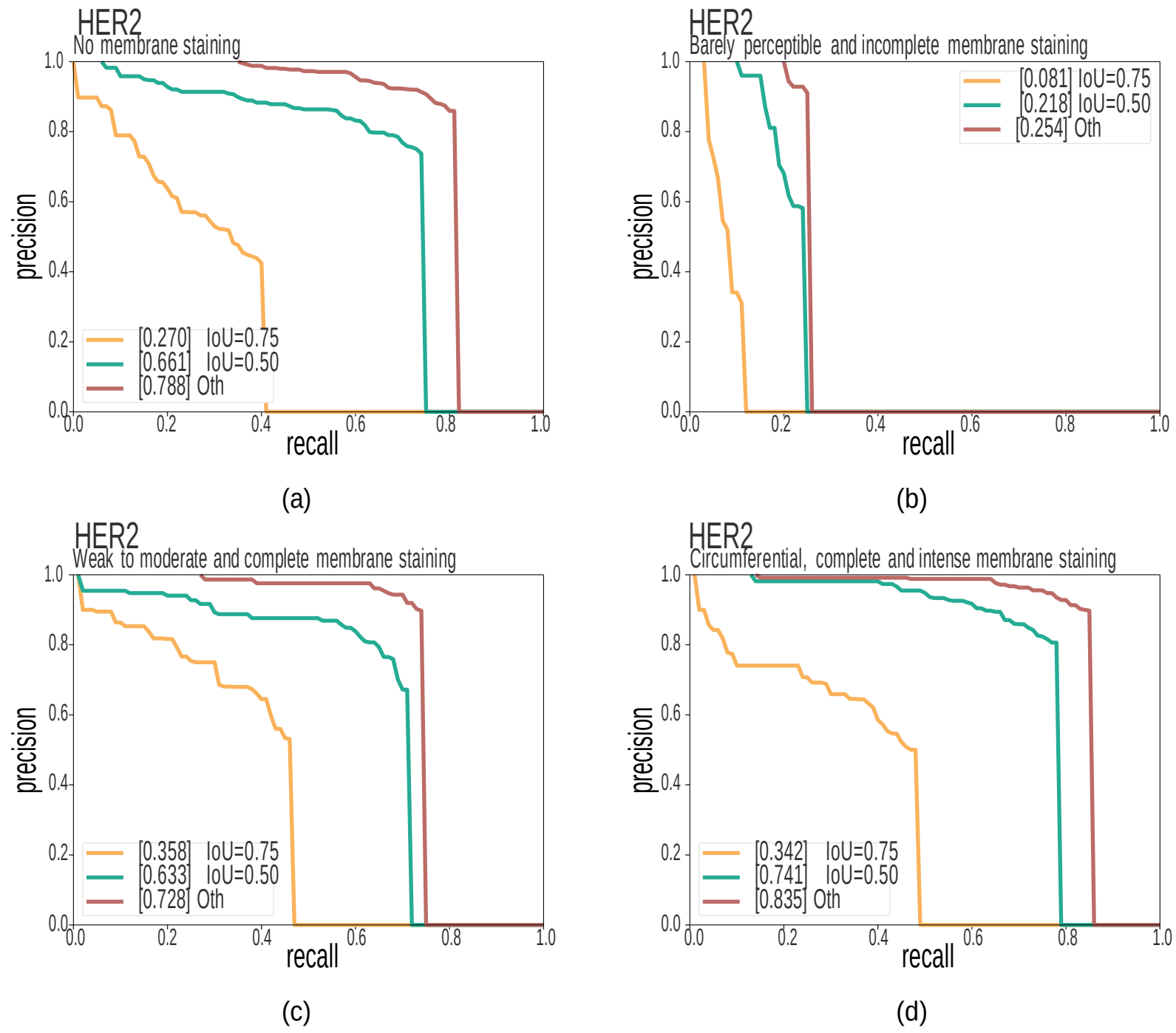

Figure 12: Precision-recall curves ($IoU$ = 0.5, $IoU$ = 0.75 and *Oth*) for no membrane staining (a), barely perceptible and incomplete membrane staining (b), weak to moderate and complete membrane staining (c) and intense and complete membrane staining (d) categories considering the test set extracted from HER2 IHC-stained images.

Finally, Fig.12 represents the precision-recall curves related to the segmentation of instances in the HER2 IHC-stained image dataset, according to the membrane appearance. The subgraphs show that for class "circumferential, intense and complete membrane staining", the size and shape of the automatically segmented nuclei more closely follows the manually segmented ones, given the distance between the $IoU$ = 0.75 and $IoU$ = 0.50 curves. Last, it should be noted that this membrane-aware segmentation model, as opposed to the one trained for quantifying nuclear biomarkers, is effective in discriminating tumour cells but shows slightly more confusion in the categorisation of the four classes considered in the HER2 IHC dataset.

### *3.2. Performance comparison with a benchmark DL-based instance segmentation model*

Having trained and validated the SOLOv2-based nuclei- and membrane-aware segmentation models, we have replicated the training and validation process on models based on Mask-RCNN, a deep learning instance segmentation algorithm that has become a reference in the field. The purpose of this comparison is to ensure that the *mAP* values obtained with the SOLOv2-based nuclei- and membrane-aware segmentation models are comparable to those resulting from applying a well-established state-of-the-art algorithm. As we did at the beginning of the previous subsection, we have optimized the $\tau_{th}$ parameter for the Mask-RCNN predictions, resulting in a value of 0.6, which will determine whether a segmented instance is considered a tumor cell.

The results are reported in Table 7 where *mAP* metrics are displayed for SOLOv2- and Mask-RCNN-based models. It can be observed that, when dealing with membrane biomarkers, the SOLOv2-based membrane-aware model outperforms Mask-RCNN in segmenting and classifying tumor cells, for all *mAP* metrics considered. However, for the trained nuclei-aware models, Mask-RCNN provides a slight improvement only in $mAP^{IoU=.50}$.

Table 7: Assessment of SOLOv2- and Mask-RCNN-based membrane-aware segmentation models applied to Ki-67, ER and PR (a) and HER2 (b) IHC-stained images considering all tumor cells.

(a)

**Nuclei-aware segmentation models**

| Instance segmentation algorithm | $mAP^{IoU=.50}$ | $mAP^{IoU=.75}$ | $mAP^{IoU=[.50:.05:.95]}$ |
|---|---|---|---|
| SOLOv2 | 0.67 | 0.35 | 0.36 |
| Mask-RCNN | 0.68 | 0.21 | 0.30 |

(b)

**Membrane-aware segmentation models**

| Instance segmentation algorithm | $mAP^{IoU=.50}$ | $mAP^{IoU=.75}$ | $mAP^{IoU=[.50:.05:.95]}$ |
|---|---|---|---|
| SOLOv2 | 0.56 | 0.26 | 0.29 |
| Mask-RCNN | 0.51 | 0.14 | 0.21 |

This performance gain, in line with those obtained in the original SOLOv2 paper and its comparison with other segmentation methods applied on the COCO dataset, support the proposed model's capacity for tumor cell segmentation and hence for the quantification and scoring of nuclear and membrane biomarkers in breast cancer immunohistochemistry images.

## 4. Conclusions and future work

This work has focused on the training and validation of an instance segmentation model based on the SOLOv2 architecture for the automatic and precise quantification of digital breast cancer immunohistochemistry images, considering both nuclear and membrane staining techniques, which has allowed the quantification of the nuclear biomarkers ER, PR and Ki-67, as well as the membrane marker HER2, all of them recommended for the diagnosis and prognosis of most cases of breast cancer.

The training of the quantification models requires establishing a methodology for the creation of datasets by expert users, in this case pathologists and technicians, in order to make the process as minimally arduous as possible. To this end, we have implemented a web-based WSI viewer with integrated annotation tools. Through this web platform, expert users created an initial set of fully manual annotations that served to train a first version of the segmentation/quantification models. In successive batches of annotations, pathologists, instead of generating the annotations from scratch, started from nuclei segmentations built using the previously trained models. The process of correcting these predictions instead of creating entirely new annotations lightened the pathologists' workload and accelerated the process of producing large training datasets.

Trained quantification models, so-called nuclei- and membrane-aware segmentation models, have been shown to perform well in terms of instance segmentation metrics extracted from the state-of-the-art. Of particular note, the trained SOLOv2-based segmentation models are able to carry out the quantification of biomarkers on immunohistochemistry breast cancer images obtained from different types of staining techniques, both nuclear and membrane. Moreover, in the special case of the nuclei-aware segmentation model, it is able to segment up to three different types of biomarkers. On the other hand, all images included in the datasets show a high tumour variability. Despite this, the images were fed directly into the deep learning models, without the need for additional pre-processing, demonstrating that the trained models are robust to changes in colour/stain or presence of artefacts in the input images. In addition to dealing with variability in shape, colour and texture of nuclei and membrane appearance, the trained nuclei- and membrane-aware segmentation models have proven to be effective in individually quantifying nuclei that appear clustered and in avoiding the segmentation of non-tumour cells, such as stromal cells or lymphocytes. Eventually, the adopted segmentation model for solving the automated quantification of immunohistochemistry images, SOLOv2, has been shown to perform better at a lower computational cost compared to a well-established benchmark model in the field of instance segmentation.

Considering the results obtained, the clinical application of these models shows potential to improve the process of immunohistochemistry image interpretation, alleviating the weaknesses that currently exist, i.e. the search for more reproducible results, automating the evaluation process in complex scoring systems and reducing the time consumption of pathologists, which would result in improved efficiency in clinical workflows as well as improved diagnostic quality, cornerstones of personalized and precision medicine.

The methods presented in this article have limitations. The training and test sets used in the study consisted of images obtained in the same laboratory, following a common staining process depending on the type of biomarker under study, and were digitised with the same scanner. Further generalisation of the results to more heterogeneous data sources could be achieved by increasing the training set of images to one with greater variability in staining procedures. Additionally, the results related to HER2 quantification show low values in terms of *mAP* for a specific class of nuclei/membrane appearance. In order to improve the results in this regard, the number of samples in the testing sets that include this casuistry should be increased.

Some future work is derived from this study. First, addressing the complete processing of WSI has become a problem, given the vast size of this type of images. An interesting approach would be to develop methods for the detection of representative hot spots for accurate biomarker quantification. Thus, the complete quantification process would include a first phase of patch selection and subsequent nuclei counting by applying the work proposed in this article. On another note, only three nuclear and one membrane biomarker were selected in the study. However, the methods could be re-trained to include other markers, such as the increasingly used nuclear marker p53 and the membrane marker E-Cadherin. Also, the study could be adapted to the casuistry present in immunohistochemical imaging of other types of cancer. Finally, the work presented would allow studies on intratumoral genetic heterogeneity, as well as correlating it with disease prognosis: by knowing the precise location and shape of tumor nuclei, it would be possible to characterize the distribution of cells in different patterns, such as clustered, mosaic-like or scattered.

In summary, results from the application of instance segmentation techniques for the quantification of nuclear and membrane markers have shown promise, with these techniques being successful in properly discerning instances even when the processed images present nuclei with different size, shape, hue levels or membrane appearance. The trained models have been implemented in an web-based platform that aids collaboration between pathologists and researchers, improving the understanding between both groups, alleviating the workload of pathologists and optimizing the obtaining of research results. Currently, this tool can be used as a decision support tool by pathologists.

## CRediT authorship contribution statement

**Blanca Maria Priego-Torres:** Conceptualization, Methodology, Software, Validation, Formal analysis, Writing - Original draft, Writing - Review editing, Visualization. **Barbara Lobato-Delgado:** Visualization, Investigation. **Lidia Atienza-Cuevas:** Investigation, Data curation, Writing - Review editing. **Daniel Sanchez-Morillo:** Formal analysis, Methodology, Investigation, Writing - Original draft, Writing - Review editing.

## Declaration of Competing Interest

The authors declare that they have no known competing financial interests or personal relationships that could have appeared to influence the work reported in this paper.

## Acknowledgments

Funding: Project PI-0032-2017. Subvención para la financiación de la investigación y la innovación biomédica y en Ciencias de la Salud en el marco de la iniciativa territorial integrada 2014–2020 para la provincia de Cádiz. Consejería de Salud. Junta de Andalucía. Unión Europea, financed by the Fondo de Desarrollo Regional (FEDER).

## References

Akakin, H. C., Kong, H., Elkins, C., Hemminger, J., Miller, B., Ming, J., Plocharczyk, E., Roth, R., Weinberg, M., Ziegler, R. et al. (2012). Automated detection of cells from immunohistochemically-stained tissues: Application to ki-67 nuclei staining. In *Medical Imaging 2012: Computer-Aided Diagnosis* (p. 831503). International Society for Optics and Photonics volume 8315.

Al-Lahham, H. Z., Alomari, R. S., Hiary, H., & Chaudhary, V. (2012). Automating proliferation rate estimation from ki-67 histology images. In *Medical Imaging 2012: Computer-Aided Diagnosis* (p. 83152A). International Society for Optics and Photonics volume 8315.

Badejo, J. A., Adetiba, E., Akinrinmade, A., & Akanle, M. B. (2018). Medical image classification with hand-designed or machine-designed texture descriptors: a performance evaluation. In *International Conference on Bioinformatics and Biomedical Engineering* (pp. 266–275). Springer.

Barricelli, B. R., Casiraghi, E., Gliozzo, J., Huber, V., Leone, B. E., Rizzi, A., & Vergani, B. (2019). ki67 nuclei detection and ki67-index estimation: a novel automatic approach based on human vision modeling. *BMC bioinformatics*, *20*, 733.

Bolya, D., Zhou, C., Xiao, F., & Lee, Y. J. (2019). Yolact: Real-time instance segmentation. In *Proceedings of the IEEE/CVF International Conference on Computer Vision* (pp. 9157–9166).

Burstein, H. J., Curigliano, G., Loibl, S., Dubsky, P., Gnant, M., Poortmans, P., Colleoni, M., Denkert, C., Piccart-Gebhart, M., Regan, M. et al. (2019). Estimating the benefits of therapy for early-stage breast cancer: the st. gallen international consensus guidelines for the primary therapy of early breast cancer 2019. *Annals of Oncology*, *30*, 1541–1557.

Calvo, J. P., Albanell, J., Rojo, F., Ciruelos, E., Aranda-López, I., Cortés, J., García-Caballero, T., Martín, M., López-García, M. Á., & Colomer, R. (2018). Consenso de la sociedad española de anatomía patológica y la sociedad española de oncología médica sobre biomarcadores en cáncer de mama. *Revista Española de Patología*, *51*, 97–109.

Chen, H., Sun, K., Tian, Z., Shen, C., Huang, Y., & Yan, Y. (2020). Blendmask: Top-down meets bottom-up for instance segmentation. In *Proceedings of the IEEE/CVF conference on computer vision and pattern recognition* (pp. 8573–8581).

Chen, L., Bao, J., Huang, Q., & Sun, H. (2019a). A robust and automated cell counting method in quantification of digital breast cancer immunohistochemistry images. *Polish Journal of Pathology*, *70*, 162–173.

Chen, X., Girshick, R., He, K., & Dollár, P. (2019b). Tensormask: A foundation for dense object segmentation. In *Proceedings of the IEEE/CVF International Conference on Computer Vision* (pp. 2061–2069).

Chlipala, E. A., Bendzinski, C. M., Dorner, C., Sartan, R., Copeland, K., Pearce, R., Doherty, F., & Bolon, B. (2020). An image analysis solution for quantification and determination of immunohistochemistry staining reproducibility. *Applied Immunohistochemistry & Molecular Morphology*, *28*, 428.

De Matos, L. L., Trufelli, D. C., De Matos, M. G. L., & da Silva Pinhal, M. A. (2010). Immunohistochemistry as an important tool in biomarkers detection and clinical practice. *Biomarker insights*, *5*, BMI–S2185.

Duffy, M., Harbeck, N., Nap, M., Molina, R., Nicolini, A., Senkus, E., & Cardoso, F. (2017). Clinical use of biomarkers in breast cancer: Updated guidelines from the european group on tumor markers (egtm). *European journal of cancer*, *75*, 284–298.

Farahani, N., Parwani, A. V., & Pantanowitz, L. (2015). Whole slide imaging in pathology: advantages, limitations, and emerging perspectives. *Pathol Lab Med Int*, *7*, 4321.

Ferlay, J., Colombet, M., Soerjomataram, I., Mathers, C., Parkin, D., Piñeros, M., Znaor, A., & Bray, F. (2019). Estimating the global cancer incidence and mortality in 2018: Globocan sources and methods. *International journal of cancer*, *144*, 1941–1953.

Garcia-Rojo, M., De Mena, D., Muriel-Cueto, P., Atienza-Cuevas, L., Dominguez-Gomez, M., & Bueno, G. (2019). New european union regulations related to whole slide image scanners and image analysis software. *Journal of pathology informatics*, *10*.

He, K., Gkioxari, G., Dollár, P., & Girshick, R. (2017). Mask r-cnn. In *Proceedings of the IEEE international conference on computer vision* (pp. 2961–2969).

He, K., Zhang, X., Ren, S., & Sun, J. (2016). Deep residual learning for image recognition. In *Proceedings of the IEEE conference on computer vision and pattern recognition* (pp. 770–778).

Hou, Y., Nitta, H., Parwani, A. V., & Li, Z. (2020). The assessment of her2 status and its clinical implication in breast cancer. *Diagnostic Histopathology*, *26*, 61–68.

Huang, P.-W., & Lai, Y.-H. (2010). Effective segmentation and classification for hcc biopsy images. *Pattern Recognition*, *43*, 1550–1563.

Irshad, H., Veillard, A., Roux, L., & Racoceanu, D. (2013). Methods for nuclei detection, segmentation, and classification in digital histopathology: a review—current status and future potential. *IEEE reviews in biomedical engineering*, *7*, 97–114.

Khameneh, F. D., Razavi, S., & Kamasak, M. (2019). Automated segmentation of cell membranes to evaluate her2 status in whole slide images using a modified deep learning network. *Computers in biology and medicine*, *110*, 164–174.

Kirkegaard, T., Edwards, J., Tovey, S., McGlynn, L., Krishna, S., Mukherjee, R., Tam, L., Munro, A., Dunne, B., & Bartlett, J. (2006). Observer variation in immunohistochemical analysis of protein expression, time for a change? *Histopathology*, *48*, 787–794.

Lee, D.-H. et al. (2013). Pseudo-label: The simple and efficient semi-supervised learning method for deep neural networks. In *Workshop on challenges in representation learning, ICML* (p. 896). volume 3.

Leung, S. C., Nielsen, T. O., Zabaglo, L. A., Arun, I., Badve, S. S., Bane, A. L., Bartlett, J. M., Borgquist, S., Chang, M. C., Dodson, A. et al. (2019). Analytical validation of a standardised scoring protocol for ki67 immunohistochemistry on breast cancer excision whole sections: an international multicentre collaboration. *Histopathology*, *75*, 225–235.

Lin, T.-Y., Dollár, P., Girshick, R., He, K., Hariharan, B., & Belongie, S. (2017). Feature pyramid networks for object detection. In *Proceedings of the IEEE conference on computer vision and pattern recognition* (pp. 2117–2125).

Lin, T.-Y., Maire, M., Belongie, S., Hays, J., Perona, P., Ramanan, D., Dollár, P., & Zitnick, C. L. (2014). Microsoft coco: Common objects in context. In *European conference on computer vision* (pp. 740–755). Springer.

López, C., Lejeune, M., Salvadó, M. T., Escrivà, P., Bosch, R., Pons, L. E., Álvaro, T., Roig, J., Cugat, X., Baucells, J. et al. (2008). Automated quantification of nuclear immunohistochemical markers with different complexity. *Histochemistry and cell biology*, *129*, 379–387.

Manni, A., Arafah, B., & Pearson, O. H. (1980). Estrogen and progesterone receptors in the prediction of response of breast cancer to endocrine therapy. *Cancer*, *46*, 2838–2841.

Markiewicz, T., Wisniewski, P., Osowski, S., Patera, J., Kozlowski, W., & Koktysz, R. (2009). Comparative analysis of methods for accurate recognition of cells through nuclei staining of ki-67 in neuroblastoma and estrogen/progesterone status staining in breast cancer. *Analytical and quantitative cytology and histology*, *31*, 49.

Narayanan, P. L., Raza, S. E. A., Dodson, A., Gusterson, B., Dowsett, M., & Yuan, Y. (2018). Deepsdcs: Dissecting cancer proliferation heterogeneity in ki67 digital whole slide images. *arXiv preprint arXiv:1806.10850*, .

Padilla, R., Passos, W. L., Dias, T. L., Netto, S. L., & da Silva, E. A. (2021). A comparative analysis of object detection metrics with a companion open-source toolkit. *Electronics*, *10*, 279.

Priego-Torres, B. M., Sanchez-Morillo, D., Fernandez-Granero, M. A., & Garcia-Rojo, M. (2020). Automatic segmentation of whole-slide h&e stained breast histopathology images using a deep convolutional neural network architecture. *Expert Systems With Applications*, *151*, 113387.

Qaiser, T., Mukherjee, A., Reddy Pb, C., Munugoti, S. D., Tallam, V., Pitkäaho, T., Lehtimäki, T., Naughton, T., Berseth, M., Pedraza, A. et al. (2018). Her 2 challenge contest: a detailed assessment of automated her 2 scoring algorithms in whole slide images of breast cancer tissues. *Histopathology*, *72*, 227–238.

Rojo, M. G., Bueno, G., & Slodkowska, J. (2009). Review of imaging solutions for integrated quantitative immunohistochemistry in the pathology daily practice. *Folia histochemica et cytobiologica*, *47*, 349–354.

Ronneberger, O., Fischer, P., & Brox, T. (2015). U-net: Convolutional networks for biomedical image segmentation. In *International Conference on Medical image computing and computer-assisted intervention* (pp. 234–241). Springer.

Saha, M., Chakraborty, C., Arun, I., Ahmed, R., & Chatterjee, S. (2017). An advanced deep learning approach for ki-67 stained hotspot detection and proliferation rate scoring for prognostic evaluation of breast cancer. *Scientific reports*, *7*, 1–14.

Sheikhzadeh, F., Ward, R. K., van Niekerk, D., & Guillaud, M. (2018). Automatic labeling of molecular biomarkers of immunohistochemistry images using fully convolutional networks. *PloS one*, *13*, e0190783.

Shu, J., Fu, H., Qiu, G., Kaye, P., & Ilyas, M. (2013). Segmenting overlapping cell nuclei in digital histopathology images. In *2013 35th Annual International Conference of the IEEE Engineering in Medicine and Biology Society (EMBC)* (pp. 5445–5448). IEEE.

Shu, J., Liu, J., Zhang, Y., Fu, H., Ilyas, M., Faraci, G., Della Mea, V., Liu, B., & Qiu, G. (2020). Marker controlled superpixel nuclei segmentation and automatic counting on immunohistochemistry staining images. *Bioinformatics*, *36*, 3225–3233.

Su, H., Yin, Z., Huh, S., Kanade, T., & Zhu, J. (2015). Interactive cell segmentation based on active and semi-supervised learning. *IEEE transactions on medical imaging*, *35*, 762–777.

Tajbakhsh, N., Jeyaseelan, L., Li, Q., Chiang, J. N., Wu, Z., & Ding, X. (2020). Embracing imperfect datasets: A review of deep learning solutions for medical image segmentation. *Medical Image Analysis*, *63*, 101693.

Vandenberghe, M. E., Scott, M. L., Scorer, P. W., Söderberg, M., Balcerzak, D., & Barker, C. (2017). Relevance of deep learning to facilitate the diagnosis of her2 status in breast cancer. *Scientific reports*, *7*, 1–11.

Wang, X., Kong, T., Shen, C., Jiang, Y., & Li, L. (2020a). Solo: Segmenting objects by locations. In *European Conference on Computer Vision* (pp. 649–665). Springer.

Wang, X., Zhang, R., Kong, T., Li, L., & Shen, C. (2020b). Solov2: Dynamic and fast instance segmentation. *Advances in Neural Information Processing Systems*, *33*.

Wang, Y., Xu, Z., Shen, H., Cheng, B., & Yang, L. (2020c). Centermask: single shot instance segmentation with point representation. In *Proceedings of the IEEE/CVF Conference on Computer Vision and Pattern Recognition* (pp. 9313–9321).

Wolff, A. C., Hammond, M. E. H., Allison, K. H., Harvey, B. E., Mangu, P. B., Bartlett, J. M., Bilous, M., Ellis, I. O., Fitzgibbons, P., Hanna, W. et al. (2018). Human epidermal growth factor receptor 2 testing in breast cancer: American society of clinical oncology/college of american pathologists clinical practice guideline focused update. *Archives of pathology & laboratory medicine*, *142*, 1364–1382.

Xing, F., Su, H., Neltner, J., & Yang, L. (2013). Automatic ki-67 counting using robust cell detection and online dictionary learning. *IEEE Transactions on Biomedical Engineering*, *61*, 859–870.

Xue, C., Deng, Q., Li, X., Dou, Q., & Heng, P.-A. (2020). Cascaded robust learning at imperfect labels for chest x-ray segmentation. In *International Conference on Medical Image Computing and Computer-Assisted Intervention* (pp. 579–588). Springer.

Xue, Y., Ray, N., Hugh, J., & Bigras, G. (2016). Cell counting by regression using convolutional neural network. In *European Conference on Computer Vision* (pp. 274–290). Springer.

Zaha, D. C. (2014). Significance of immunohistochemistry in breast cancer. *World journal of clinical oncology*, *5*, 382.

Zarella, M. D., Bowman, D., Aeffner, F., Farahani, N., Xthona, A., Absar, S. F., Parwani, A., Bui, M., & Hartman, D. J. (2019). A practical guide to whole slide imaging: a white paper from the digital pathology association. *Archives of pathology & laboratory medicine*, *143*, 222–234.

Zhang, L., Lin, L., Liang, X., & He, K. (2016). Is faster r-cnn doing well for pedestrian detection? In *European conference on computer vision* (pp. 443–457). Springer.

Zhang, R., Tian, Z., Shen, C., You, M., & Yan, Y. (2020a). Mask encoding for single shot instance segmentation. In *Proceedings of the IEEE/CVF Conference on Computer Vision and Pattern Recognition* (pp. 10226–10235).

Zhang, X., Cornish, T. C., Yang, L., Bennett, T. D., Ghosh, D., & Xing, F. (2020b). Generative adversarial domain adaptation for nucleus quantification in images of tissue immunohistochemically stained for ki-67. *JCO Clinical Cancer Informatics*, *4*, 666–679.